\documentclass[times]{elsarticle}

\usepackage{bsymb}

\usepackage{cite}
\usepackage{amsmath,amssymb,amsfonts}
\usepackage{relsize}
\usepackage{algorithmic}
\usepackage{graphicx}
\usepackage{subfig}

\usepackage{textcomp}
\def\BibTeX{{\rm B\kern-.05em{\sc i\kern-.025em b}\kern-.08em
    T\kern-.1667em\lower.7ex\hbox{E}\kern-.125emX}}

\usepackage{mathtools}

\usepackage[utf8]{inputenc}
\usepackage[T1]{fontenc}
\usepackage{microtype}

\usepackage{color}
\definecolor{gray}{rgb}{0.4,0.4,0.4}
\definecolor{darkblue}{rgb}{0.0,0.0,0.6}
\definecolor{cyan}{rgb}{0.0,0.6,0.6}
\definecolor{keycolor}{rgb}{0,0,0.8}     
\definecolor{labelcolor}{rgb}{0,0.4,0.8} 
\definecolor{codecolor}{rgb}{0,0,0}      
\definecolor{inhcolor}{rgb}{0.6,0.2,0}   
\definecolor{cmtcolor}{rgb}{0,0.4,0}     

\usepackage{bsymb}

\usepackage{listings}

\usepackage{color}
\definecolor{gray}{rgb}{0.4,0.4,0.4}
\definecolor{darkblue}{rgb}{0.0,0.0,1.0}
\definecolor{cyan}{rgb}{0.0,0.6,0.6}

\lstdefinelanguage{MPS}
{
  morestring=[b]",
  morestring=[s]
  ,
  morecomment=[s]{/*}{*/},
  stringstyle=\color{black},
  identifierstyle=\color{black},
  keywordstyle=\color{darkblue},
  morekeywords={Natural Language, Query Engine, Relational Databases, Generative AI, SQL}
  ,
  otherkeywords = {=,&,(,),\{,\},>,<,",:,?},
}

\usepackage{lineno,hyperref}
\modulolinenumbers[5]

\journal{Arxiv}

\usepackage{breqn}
\usepackage{longtable}

\usepackage[linewidth=1pt]{mdframed}
\usepackage{multicol}

\usepackage{flushend}

\begin{document}

\begin{frontmatter}

\title{Natural Language Query Engine for Relational Databases using Generative AI}

\author{Steve Tueno}
\address{IBM France}
\ead{steve.tueno.fotso@ibm.com}


\begin{abstract}

The growing reliance on data-driven decision-making highlights the need for more intuitive ways to access and analyze information stored in relational databases. However, the requirement of SQL knowledge has long been a significant barrier for non-technical users. This article introduces an innovative solution that leverages \textbf{generative AI} to bridge this gap, enabling users to \textbf{query databases using natural language}. Our approach automatically translates natural language queries into SQL, \textit{ensuring both syntactic and semantic correctness}, while also generating clear, natural language responses from the retrieved data. By streamlining the interaction between users and databases, this method empowers individuals without technical expertise to engage with data directly and efficiently, democratizing access to valuable insights and enhancing productivity.

\end{abstract}


\end{frontmatter}


\section*{Introduction}

The ability to query relational databases using natural language has been a long-standing challenge, particularly for non-expert users who lack the technical knowledge of SQL. Over the years, numerous efforts have been made to bridge the gap between natural language and structured database queries, aiming to democratize access to data and make querying more intuitive. However, existing solutions, whether patented technologies, research papers, or platform-specific implementations, fall short in several key areas. They often focus exclusively on translating natural language into SQL without ensuring that the generated queries are syntactically and semantically correct and all overlook the need for response generation in well-formed natural language \citep{US20180181613A1} \citep{US11714841B2} \citep{CN115048407A} \citep{US10592505B2} \citep{US10657125B1} \citep{US5924089A} \citep{CN109918453B} \citep{US9652451B2}. Furthermore, many fail to account for the complex business rules and domain-specific constraints that are essential for accurate data retrieval in real-world scenarios \citep{CN117112732A} \citep{US20190197185A1}  \citep{US20180181613A1} \citep{US11714841B2} \citep{CN115048407A} \citep{US10592505B2} \citep{US10657125B1} \citep{US5924089A} \citep{CN109918453B}
\citep{SQLizer} \citep{CatSQL} \citep{DIY} \citep{transforming_nl_text_to_sql} \citep{dataiku_llm_nl_querying} \citep{database_queries_llm_langchain} \citep{nl_execute_sql_queries}.

Patented methods such as \citep{US11714841B2} emphasize the translation of natural language into SQL but overlook the need for query validation and response generation in well-formed natural language. Similarly, research efforts like \citep{CatSQL} and blog articles on platforms such as \citep{dataiku_llm_nl_querying} and \citep{nl_execute_sql_queries} demonstrate improvements in natural language querying but struggle with scalability, generalizability, and the integration of complex business rules. These shortcomings highlight the need for a more robust solution that can handle both the technical aspects of SQL generation and the contextual understanding required to apply business rules effectively.

This article introduces a novel approach that addresses these challenges by leveraging generative AI to enhance the entire querying process. Our method not only translates natural language into SQL but also ensures the correctness of the generated queries through multi-step validation. It incorporates business rules using vector database technologies, enabling the system to handle complex queries with accuracy and generate natural language responses that are clear and contextually appropriate. By doing so, it provides a scalable, user-friendly solution for querying relational databases, empowering non-expert users to interact with data intuitively and confidently.

In the following sections, we will explore the drawbacks of existing solutions in greater detail and demonstrate how our contribution overcomes these limitations to deliver a comprehensive and reliable natural language interface for database querying.

\section{Background}

To fully understand the contribution presented in this article, it is important to grasp several foundational technical concepts. 

\subsection{Relational Databases}

A \textbf{relational database} is a type of database that stores data in tables (or relations), where each table consists of rows and columns. Each row represents a record, and each column represents an attribute of the record. These tables are structured in a way that relationships between different types of data can be defined and maintained.

For example, a relational database for an e-commerce platform might have separate tables for customers, orders, and products. Relationships between these tables, such as which customer placed which order, are defined using \textbf{primary keys} (unique identifiers for each record) and \textbf{foreign keys} (references to a primary key in another table).

Relational databases are widely used because they:
\begin{itemize}
\item[•] Offer structured and organized data storage
\item[•] Support complex queries across multiple tables
\item[•] Ensure data integrity through constraints and relationships
\item[•] Provide powerful querying capabilities using SQL (Structured Query Language)
\end{itemize}

Popular relational database management systems (RDBMS) include MySQL, PostgreSQL, IBM DB2, Microsoft SQL Server, and Oracle Database.

\subsection{Structured Query Language (SQL)}

\textbf{SQL (Structured Query Language)} is the standard language used to interact with relational databases. It allows users to perform various operations such as retrieving, inserting, updating, and deleting data within the database.

\textbf{Key components of SQL:}
\begin{itemize}
\item[•] \textbf{SELECT}: Used to query and retrieve specific data from the database.
\item[•] \textbf{INSERT}: Adds new data into a table
\item[•] \textbf{UPDATE}: Modifies existing data in a table.
\item[•] \textbf{DELETE}: Removes data from a table.
\end{itemize}

Eg: 
\begin{lstlisting}
SELECT * FROM tpch.tiny.customer WHERE order_amount>0
\end{lstlisting}

SQL supports complex queries that can join multiple tables and perform data aggregation, filtering, and sorting. However, constructing SQL queries can be a challenge for non-technical users who lack familiarity with the database structure or SQL syntax. This is one of the primary barriers that this article’s contribution seeks to overcome.

\subsection{Large Language Models (LLMs)}

\textbf{Large Language Models (LLMs)} are a type of artificial intelligence model that are trained on vast amounts of textual data to understand and generate human language. They are built on deep learning architectures, such as transformers, and are capable of performing a wide variety of natural language processing tasks, including:
\begin{itemize}
\item[•] Language translation
\item[•] Text generation
\item[•] Sentiment analysis
\item[•] Question answering
\end{itemize}

LLMs, like \textbf{\textit{Mixtral}} \citep{mixtral}, are especially adept at understanding context and generating coherent, contextually relevant responses. In the context of querying databases, LLMs can be used to interpret natural language input, generate SQL queries, and even produce responses in well-formed natural language based on the retrieved data.

LLMs are trained on diverse datasets and can generalize across different domains, making them highly valuable for applications like natural language interfaces for databases. However, while LLMs can generate SQL queries, their output needs validation to ensure that it is both syntactically and semantically correct, a gap that this contribution addresses.

\subsection{Vector Databases}

A \textbf{vector database} \citep{vector_databases} is a type of database optimized for storing and querying high-dimensional vectors, which are numerical representations of data. These vectors are typically generated by machine learning models, such as embedding models used in natural language processing or image recognition. 

Vectors capture the semantics of data, meaning they represent the meaning or context of the information in a way that allows for similarity searches. For example, vectors can represent words, sentences, or even entire documents, and similar data points will have vectors that are close to each other in the vector space.

In the context of this contribution, a vector database is used to:
\begin{itemize}
\item[•] \textbf{Store embeddings} of database structures (tables, columns, possible values of columns, etc.) and domain specific knowledge/rules.
\item[•] \textbf{Retrieve relevant information} based on the semantic similarity of a user’s natural language query to stored data.
\end{itemize}
  
This allows the system to apply business rules more effectively and ensure that the generated SQL queries are contextually appropriate. Vector databases, such as \textbf{Pinecone}, \textbf{Chroma DB}, and \textbf{Milvus} (distributed in \textbf{IBM watsonx.data} \citep{wxdata}), are increasingly used in applications requiring fast and scalable similarity searches.

\subsection{IBM watsonx.data}
\textbf{IBM watsonx.data} \citep{wxdata} is a modern data management and lakehouse platform designed to handle large-scale structured and unstructured data. It offers flexible, high-performance data management capabilities, making it ideal for organizations that need to store and analyze vast amounts of information. Watsonx.data is part of IBM’s watsonx AI and data platform \citep{wx}, which integrates both classical and generative artificial intelligence tools with enterprise-grade data solutions.

To prototype the described approach, watsonx.data has been used both as a relational database (through its presto query engine) and as a vector store (storing vectorized representations of database schemas and business rules via Milvus). 

\subsection{IBM watsonx.ai}

\textbf{IBM watsonx.ai} \citep{wxai} is part of the IBM watsonx AI and data platform \citep{wx}, bringing together new generative AI capabilities powered by foundation models and traditional machine learning into a powerful studio spanning the full AI lifecycle. It helps businesses leverage the power of large language models, machine learning workflows, and data analytics to drive innovation.

In the context of the prototype of the proposed natural language querying method, watsonx.ai plays a pivotal role by providing access to large language models that can interpret natural language input, generate SQL queries, validate results and build user friendly answers. 
watsonx.ai provides access to pre-trained and fine-tuned large language models, including cutting-edge models such as LLama 3 and Mixtral, which are used in the prototype.

\subsection{Business Rules and Domain-Specific Knowledge}

\textbf{Business rules} are constraints or logic that govern the way an organization operates and how data should be handled. These rules define relationships, computations, or conditions that apply to specific business contexts.

For example, in a financial database, business rules might specify that the "total sales" should be calculated by summing the product of quantity and price for each sale, or that certain products can only be sold in specific regions due to regulatory constraints. These rules are often domain-specific, meaning they vary depending on the industry or organization.

When querying databases, business rules must be considered to ensure that the results accurately reflect the real-world operations and constraints. This contribution uses vector databases to store business rules and integrates them into the query generation and validation process, ensuring that the SQL queries respect these rules.

\section{Existing Solutions and Their Drawbacks}

Over the years, numerous attempts have been made to bridge the gap between natural language and SQL querying, aiming to make databases more accessible to non-expert users. Despite some progress, current solutions often fall short in several critical areas, particularly in handling the complexity of business rules and ensuring the correctness of SQL queries. This section reviews some notable patents, publications, and products, highlighting their contributions and limitations.

\subsection{Overview of existing products, publications, and patents}

\citep{US11714841B2} introduces a method for translating natural language queries into SQL. The system focuses on parsing the user's input and converting it into SQL syntax. While this approach simplifies the SQL generation process for non-technical users, it has a significant limitation: it does not verify whether the generated SQL query is syntactically and semantically correct. This omission can lead to incorrect query results or SQL errors when queries are executed. Additionally, the patent does not provide a method for converting the results back into a well-formed natural language response, leaving users to interpret raw query outputs, which can be cumbersome for non-experts.

\citep{CN115048407A} \citep{US10592505B2} \citep{US10657125B1} \citep{US5924089A} \citep{CN109918453B} \citep{US9652451B2} describe various methods for natural language to SQL or structured query translation. While they make strides in automating the SQL generation process, they lack a robust mechanism for handling the complex business rules and domain-specific requirements that are common in real-world applications. This can result in SQL queries that are technically valid but semantically incorrect, failing to capture the full context or intent of the user’s question. Furthermore, they do not offer a well-formed natural language response, leaving users with raw SQL to interpret on their own.

\citep{DIY} and \citep{CN117271557A} claim to use AI to provide a user-friendly explanation of SQL queries generated by a black box AI system. Their aim is to provide the end users with a way to manually ensure the correctness of a query based on high level explanations of what the query is supposed to achieve. \citep{CN117271557A} provide the user-friendly explanations based on some business rules captured through a knowledge graph. Using the business rules, they claim to produce a procedural explanation of a SQL Query convenient for users. 

\citep{CatSQL} is a research paper that proposes a hybrid framework combining rule-based methods with deep learning techniques to translate natural language queries into SQL. This approach improves accuracy in generating SQL queries by leveraging domain-specific rules and machine learning models. However, despite its advancements, CatSQL still faces challenges when handling complex and ambiguous queries because it does not consider business rules and domain-specific requirements that are common in real-world applications. This can often lead to incorrect results, particularly in business contexts with intricate rules and relationships. In addition, the approach does not effectively address the semantic correctness of the SQL queries it generates.  Additionally, like the other approaches, CatSQL does not generate natural language responses, requiring users to interpret raw SQL queries.

\citep{dataiku_llm_nl_querying} describes the use of Large Language Models (LLMs) to perform natural language querying in the Dataiku platform. While the integration of LLMs into the querying process represents a step forward, the solution is largely platform-dependent, limiting its scalability and generalization across other systems. Furthermore, the method lacks the depth to handle  domain specific knowledge and business rules. The generated SQL queries are also not subject to rigorous syntactic or semantic validation, which could result in inaccurate results. 

\citep{database_queries_llm_langchain} explores the use of LangChain and LLMs to generate SQL queries from natural language input. While it offers a functional method for query generation, its scope remains limited to relatively straightforward queries and basic use cases. Complex database structures and intricate business logic are not adequately addressed, limiting its applicability in enterprise environments with more sophisticated needs. Like other existing methods, it neither offer semantic validation for SQL queries, nor natural language responses, which limit its usability by business users. Same drawbacks for \citep{nl_execute_sql_queries}.

\subsection{Synthesis of common drawbacks}

\begin{itemize}
\item[•] \textbf{Lack of SQL Query Validation}: None of the reference uses AI to enable the automatic verification and refinement of each generated SQL query, along with data retrieved from the database, to provide user with a semantically correct natural language response.

\item[•] \textbf{Handling Complex Business Rules}: Existing methods often struggle to incorporate complex business rules, which are essential for ensuring that the queries align with the specific requirements and constraints of the business domain. 

\item[•] \textbf{Scalability and Generalization}: Most of the existing solutions are rule-based approaches that may not generalize well across different domains.

\item[•] \textbf{User-Friendliness}: The solutions require users to have a basic understanding of SQL or the underlying database schema, since the output it either a SQL query (most often), or the raw output from executing the query, which can be a barrier for non-expert users.

\end{itemize}

\section{Proposed Method: A Natural Language Assistant for Relational Databases}

\begin{figure*}

 \begin{center}
 \includegraphics[width=1.3\linewidth]{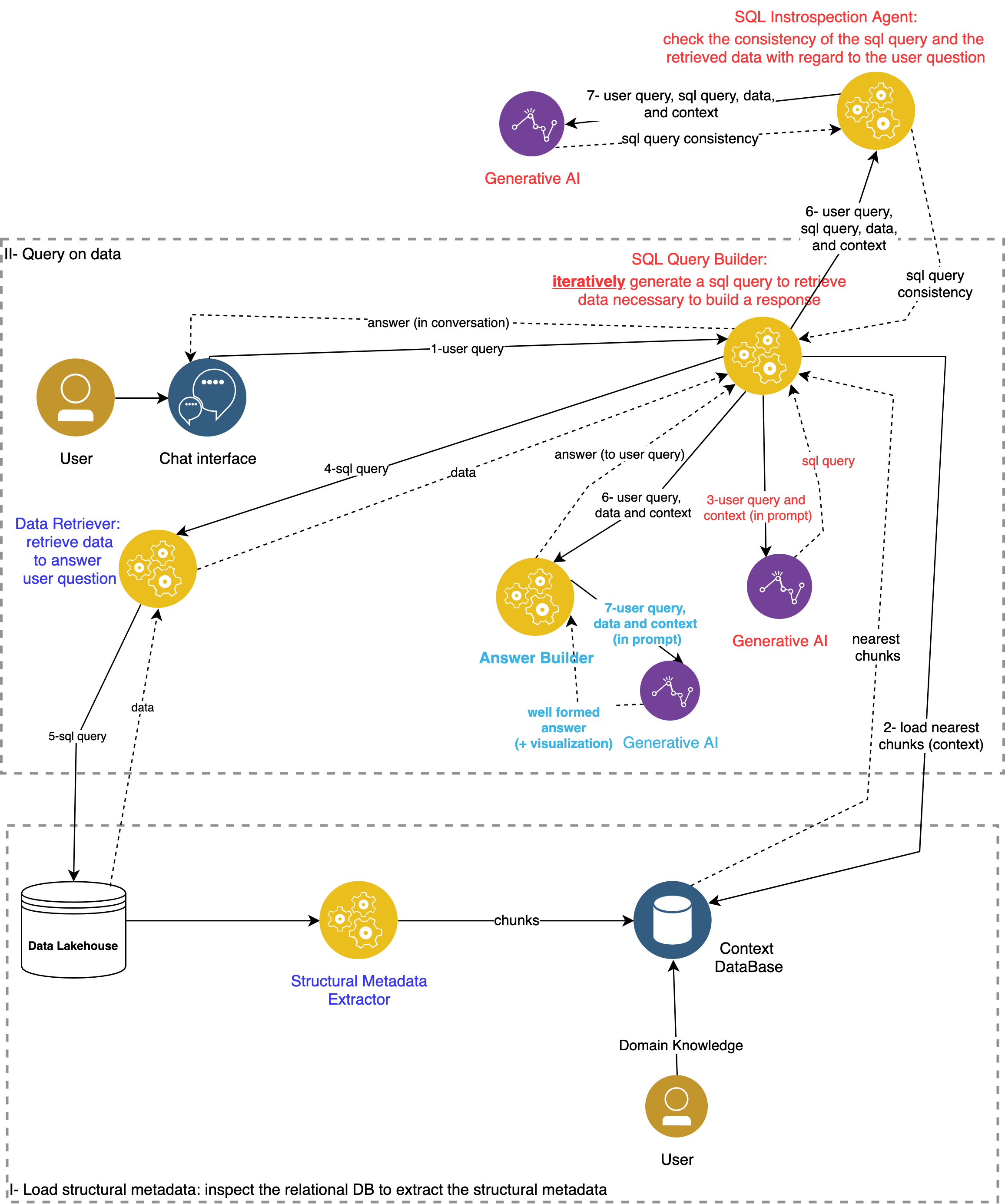}\caption{\label{approach_overview} Architecture of the Natural Language Assistant for Relational Databases} 
 \end{center}
 
 \end{figure*}

The proposed method \ref{approach_overview}  introduces a robust and intuitive way for non-technical users to interact with relational databases using natural language. By leveraging \textbf{Generative AI} to translate natural language queries into SQL, the system ensures both syntactic and semantic correctness, while also integrating \textbf{business rules} and \textbf{domain-specific knowledge}. This section provides a detailed description of each component of the method and explains how they work together to form an end-to-end solution for querying relational databases.

\subsection{Overview of the Method}

The system is designed to handle a wide range of queries, whether they relate to the structure of the database or require the retrieval of data. The process can be broken down into five key stages:
\begin{itemize}
\item \textbf{Database Structure Scanning}
\item \textbf{Business Rules Integration}
\item \textbf{Natural Language Query Processing}
\item \textbf{SQL Query Generation and Validation}
\item \textbf{Iterative Query Refinement and Natural Language Response Generation}
\end{itemize}

Each of these stages plays a critical role in transforming user input into accurate, meaningful results while minimizing the need for technical expertise.

\subsection{Database Structure Scanning}

The first step of the proposed method is the \textbf{automatic database structure scanning}, which involves analyzing the relational database to understand the schema and data distribution. This step is essential for building an accurate representation of the database that can be used to generate meaningful SQL queries.

Key elements scanned and stored include:
\begin{itemize}
\item[•] \textbf{Table structure}: The system scans the names, relationships, and primary/foreign keys of all tables in the database. Primary and foreign key relationships between tables are captured, ensuring that queries involving multiple tables can be handled accurately.
\item[•] \textbf{Column information}: For each table, the system retrieves the list of columns and their data types (e.g., integer, string, date).
\item[•] \textbf{Categorical values}: For categorical columns (those containing predefined sets of values like "region" or "status"), the system collects the possible values.
\end{itemize}

Once this information is gathered, it is \textbf{vectorized} using a text embedding model and stored in a \textbf{vector database}. This process encodes the structural elements of the database into high-dimensional vectors, which can be quickly searched and retrieved based on the user’s natural language query.

\subsection{Business Rules Integration}

Real-world databases and real-world business domains often come with complex \textbf{business rules} and \textbf{domain-specific knowledge} that must be considered and often applied during data retrieval. The proposed method integrates these rules directly into the query generation process by storing them in the vector database alongside the database structure.

Key aspects of this integration include:
\begin{itemize}
\item[•] \textbf{Storing domain-specific rules}: Business rules, such as constraints on data usage or logic for calculating specific metrics, are vectorized using a text embedding model and stored in the vector database. For example, a rule might dictate that "total revenue" should always be calculated as "quantity × price" rather than simply summing sales values.
\item[•] \textbf{Semantic filtering}: The system uses \textbf{semantic search} to filter relevant business rules based on the natural language query. By comparing the user’s query to the vectorized rules, the system can ensure that the appropriate rules are applied when generating SQL queries.
\item[•] \textbf{Dynamic rule updates}: Business rules evolve over time. The system allows for dynamic updates of the vectorized rules so that new constraints or relationships can be incorporated into future queries.
\end{itemize}

\subsection{Natural Language Query Processing}

When a user submits a natural language query, the system processes it to understand the \textbf{intent} and \textbf{context} of the question. This involves multiple stages of natural language understanding and query intent classification.

\begin{itemize}
\item[•] \textbf{Intent classification}: The system first determines whether the user’s query pertains to the structure of the database (e.g., "What tables are available?") or requires the retrieval of data (e.g., "Show me all orders from last year").
\item[•] \textbf{Context understanding}: Using a combination of vectorization of the natural language query using a text embedding model and retrieval from the vector database, the system interprets the meaning behind the user’s question. This involves matching the user’s query to relevant parts of the database structure and business rules stored in the vector database.
\item[•] \textbf{Relevant context retrieval}:  the system retrieves the most relevant text representations of the database structure and business rules. This ensures that the SQL query will reflect the necessary relationships, constraints, and logic.
\end{itemize}

\subsection{SQL Query Generation and Validation}

Once the system understands the user’s intent and retrieves the relevant information, it generates a SQL query using a large language model. This step is crucial because it automates the process of transforming natural language into executable SQL code without requiring the user to have any technical knowledge of SQL.

The system uses LLMs to generate SQL queries based on the retrieved structural information and business rules. For example, if the user asks for "total sales by region for last year," the system generates a SQL query that joins the relevant tables and applies the necessary filters and computations. During SQL generation, the system ensures that business rules, such as special calculations or domain-specific constraints, are applied to the query. This avoids common pitfalls where queries might technically run but return incorrect or incomplete data due to missing context.

To ensure accuracy and reliability, the system includes an automatic two-step validation process:

\begin{itemize}
\item \textbf{Syntactic validation}: The generated SQL is first checked for \textbf{syntactic correctness} by executing the query while limiting the number or returned rows. If any errors occur (e.g., syntax errors or invalid references), the system detects them and enters an iterative refinement process.
\item \textbf{Semantic validation} (aka \textbf{Introspection}): If the SQL query passes syntactic validation, it is subjected to \textbf{semantic validation}. This step uses \textbf{Generative AI} to evaluate the query’s logic, ensuring that it aligns with business rules and returns data that correctly answers the user’s question. If any semantic issues are identified, the system refines the query.
\end{itemize}

This dual-layer validation ensures that the queries generated by the system are both technically correct and semantically meaningful, delivering accurate results.

\subsection{Iterative Query Refinement and Natural Language Response Generation}

If the validation process detects errors, either syntactic or semantic, the system engages in \textbf{iterative refinement}. Using \textbf{Generative AI}, the system makes necessary adjustments to the SQL query based on the identified issues, such as correcting joins between tables, fixing column references, or adjusting the logic to comply with business rules.

Once a valid SQL query has been generated and validated, the system retrieves the full dataset from the database required to build an answer to the user question. The final step involves generating a \textbf{well-formed natural language response} based on the retrieved data, which is then presented to the user (see \ref{preview_structure} and \ref{preview_data}): using \textbf{Generative AI}, the system converts the data into an easy-to-understand response. For example, if the SQL query returns sales data, the system might generate a response like, \textit{"The total sales for last year were \$5 million, with the highest sales coming from the Western region."}.

\subsection{User feedback loop}

Users can ask follow-up questions, refine their queries if the response does not fully meet their needs, or provide additional business rules to consider, initiating further iterations of the process.

By iteratively refining the SQL query and generating a natural language response, the system ensures that the user receives clear and accurate answers without needing to interact with raw SQL or database outputs.

\section{Comparison with Existing Solutions}

\begin{figure*}

 \begin{center}
 \includegraphics[width=1.3\linewidth]{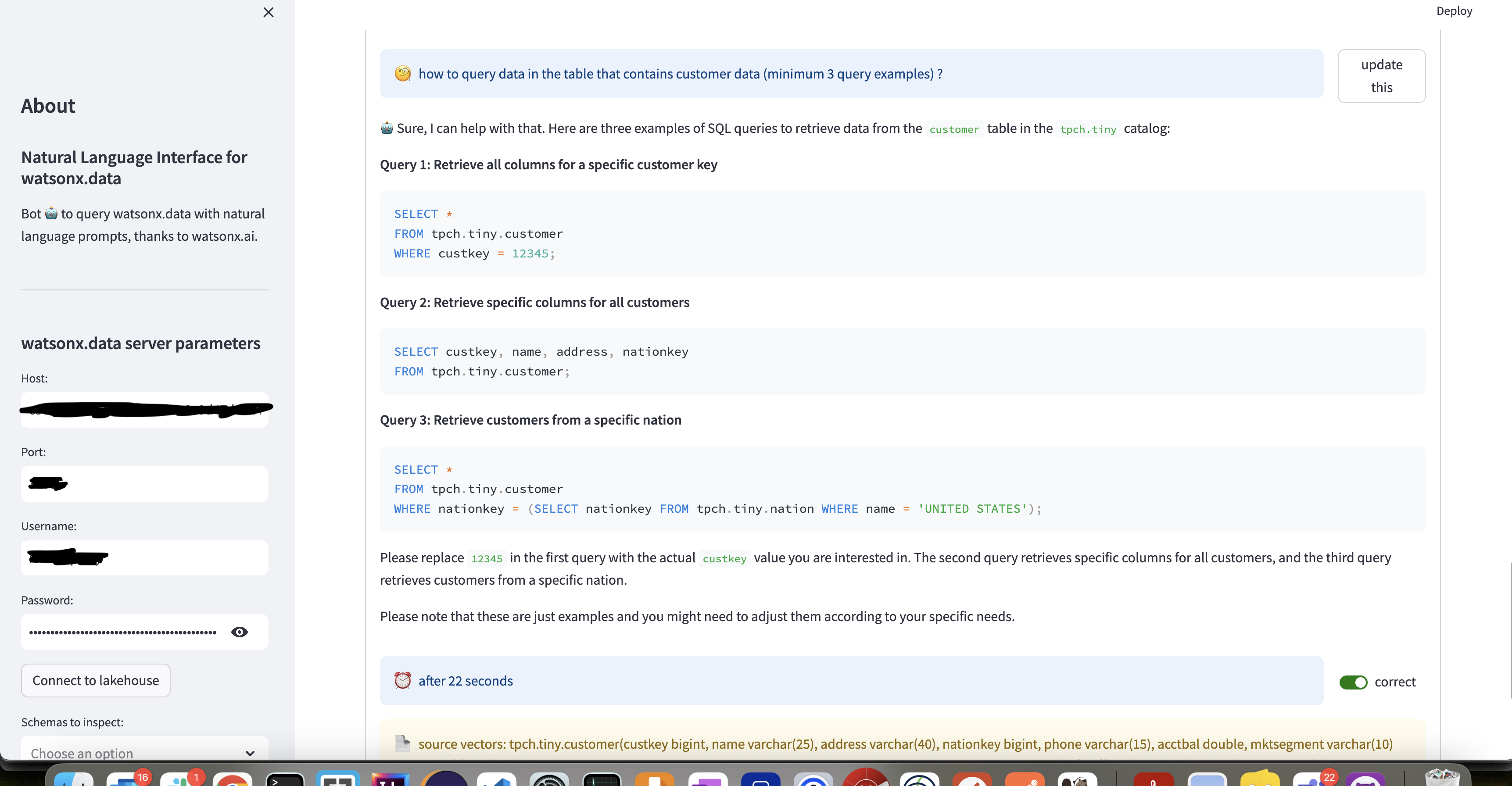}\caption{\label{preview_structure} Preview of a prototype of the approach: query on structure} 
 \end{center}
 
 \end{figure*}
 
 \begin{figure*}

 \begin{center}
 \includegraphics[width=1.3\linewidth]{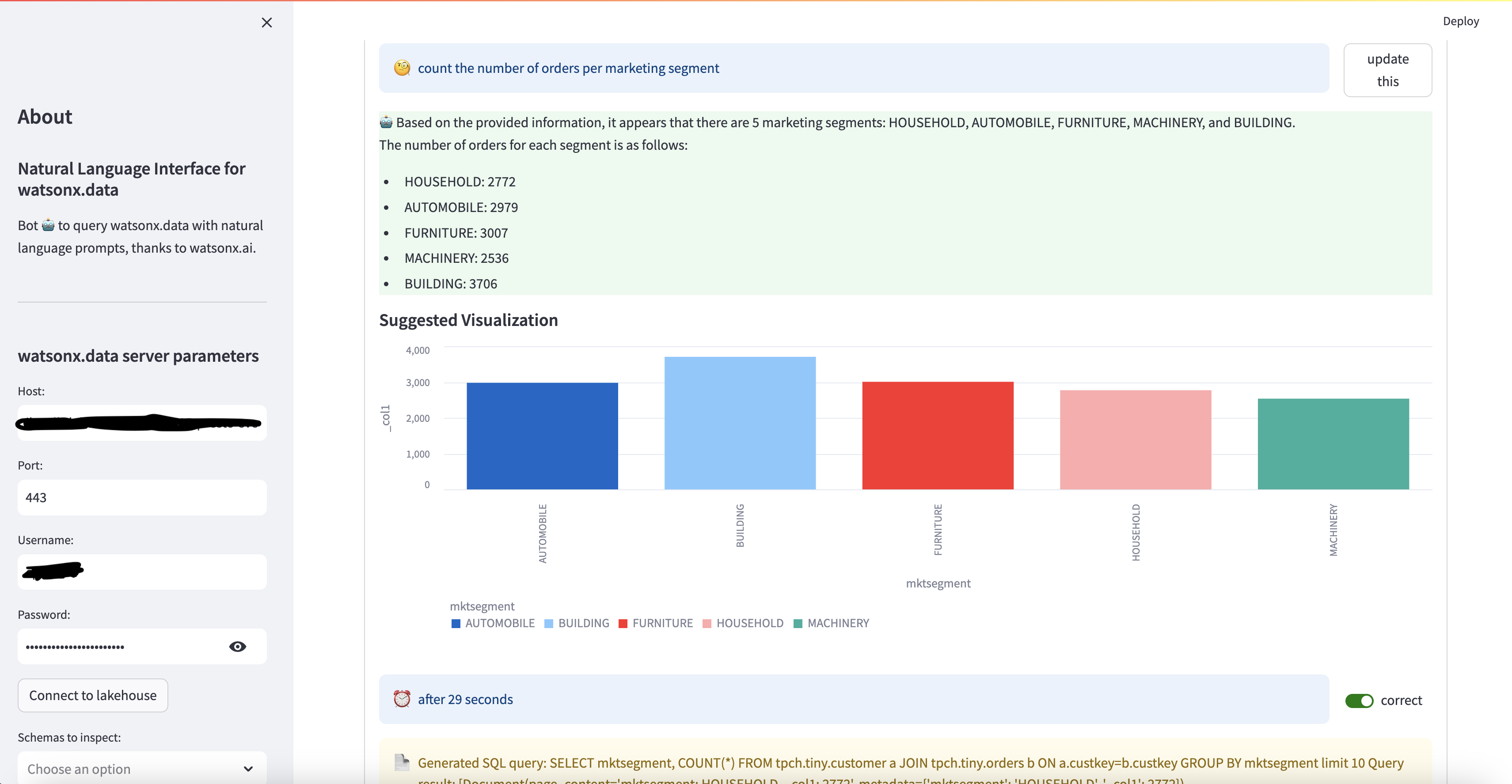}\caption{\label{preview_data} Preview of a prototype of the approach: query on data} 
 \end{center}
 
 \end{figure*}

The proposed method offers several key advantages:

\begin{itemize}
\item[•] \textbf{Accessibility}: Non-technical users can interact with databases using natural language, eliminating the need to learn SQL or understand database schemas.
\item[•] \textbf{Syntactic and Semantic Correctness}: The approach ensures that generated SQL queries are both syntactically valid and semantically aligned with business rules and user intent.
\item[•] \textbf{Business Rule Integration}: Complex domain-specific knowledge is embedded directly into the query process, ensuring accurate and context-aware results.
\item[•] \textbf{Iterative Refinement}: The system automatically corrects and refines queries to handle errors, reducing the need for manual intervention.
\item[•] \textbf{Natural Language Response}: Users receive well-formed, human-readable responses instead of raw SQL query outputs, enhancing user experience and understanding. The response can be textual, tabular or graphical (see \ref{preview_structure} and \ref{preview_data}).

\end{itemize}

Going forward from \citep{CN117271557A} which uses business knowledge to provide explanations to the user regarding a SQL code generated by a black box AI, and from \citep{CN117112732A} \citep{US20190197185A1}  which uses patterns of queries and probabilistic alignment between the natural language question and the patterns to produce a SQL query from the natural language question (which limits its generalizability), our approach relies on Generative models for SQL query generation, which are powerful and much more flexible Natural Language Processing artifacts. It makes use of Generative AI to iteratively and automatically verify and refine each generated SQL query, along with data retrieved from the database, to provide user with a semantically correct natural language response. Thus, no SQL query understanding is expected from the end user to deal with the system. 

In addition, going forward from \citep{CN117271557A} where it is not disclosed whether the business knowledge is considered by the black box AI when generating the SQL code, and from \citep{CN117112732A} \citep{US20190197185A1},  our approach stores information about the structure and business rules in a vector database which greatly accelerates and optimizes identification of relevant context information to consider for the generation of the SQL query when provided with a natural language user question. That knowledge is considered when generating SQL query to ensure semantic correctness of results.

\section{Prototyping and Evaluation}

\begin{figure*}

 \begin{center}
 \includegraphics[width=1.3\linewidth]{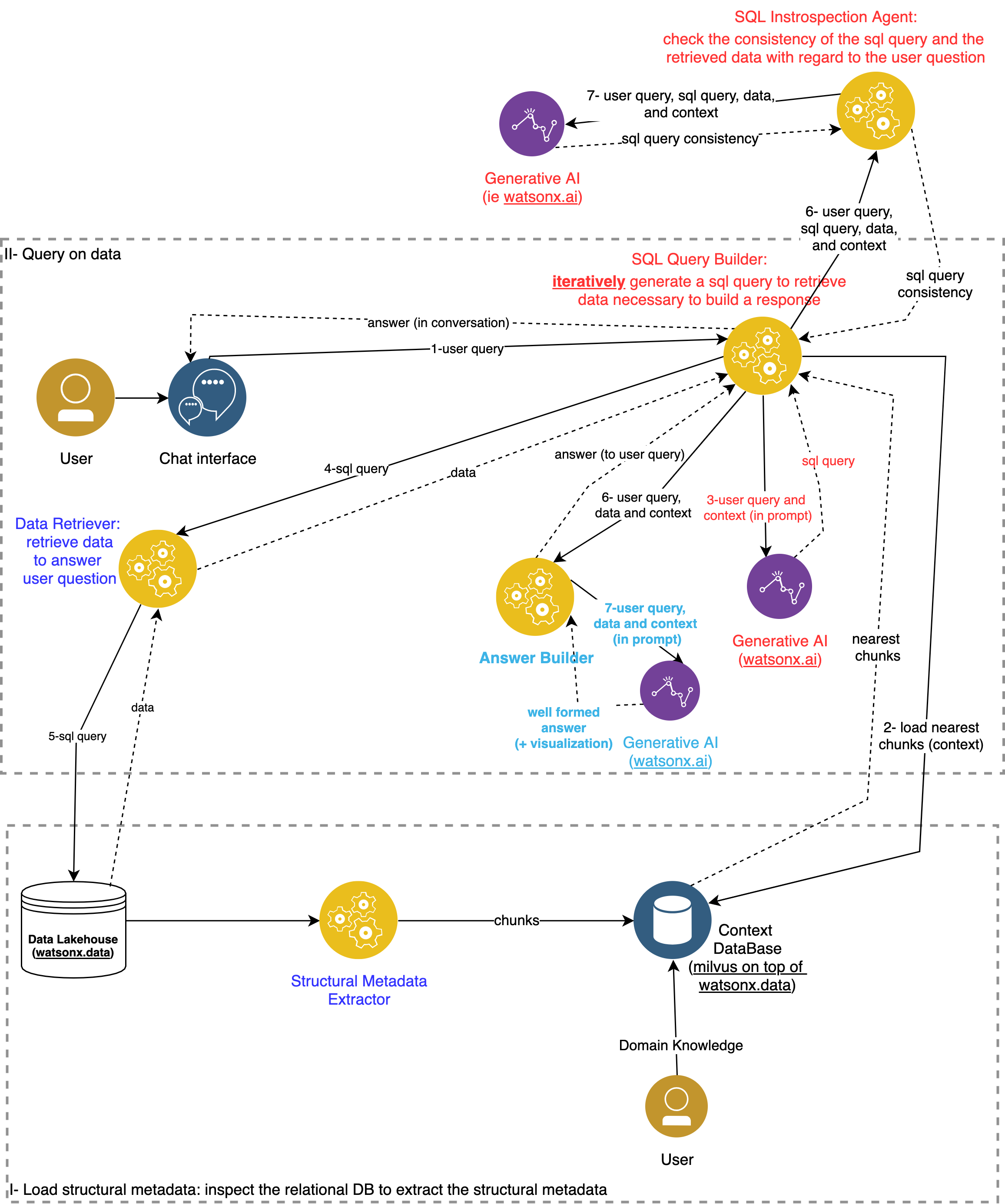}\caption{\label{prototype_overview} Architecture of the Prototype of the Natural Language Assistant for Relational Databases} 
 \end{center}
 
 \end{figure*}

To validate the proposed method for natural language querying of relational databases, a prototype was developed using a suite of IBM technologies (see \ref{prototype_overview} for architecture) and evaluated against the well-known BIRD-SQL benchmark \citep{li2024can}. This section details the tools and technologies employed in building the prototype, the evaluation methodology, and the initial results that demonstrate the effectiveness of the approach.

\subsection{Technology Stack}

The prototype was implemented using a combination of state-of-the-art tools to handle different aspects of the solution, including vector and relational databases, and large language models. The main components of the technology stack are as follows:

\begin{itemize}
\item[•] \textbf{Milvus on top of IBM watsonx.data}: Milvus was used as the vector database to handle storage of the vectorized representations of relational database structural metadata and business rules. 
\item[•] \textbf{IBM watsonx.data}: IBM watsonx.data served as the relational database component, leveraging presto, its SQL analytics engine.
\item[•] \textbf{IBM watsonx.ai}: The generative AI models used in this prototype were integrated through watsonx.ai, a platform designed for large-scale AI workloads. Specifically, the prototype utilized models such as \textbf{LLama 3}, primarily used for natural language understanding and SQL generation and validation, and \textbf{Mixtral} for generation of natural language responses based on the retrieved data. 
\item[•] \textbf{Python for Orchestration}: The entire workflow was orchestrated using a Python backend and a streamlit frontend, which served as the backbone for connecting the different components, ranging from handling user input, managing interactions with Milvus and watsonx.data, invoking LLMs through watsonx.ai, and iterating through the validation and refinement processes.
\end{itemize}

\subsection{Evaluation Methodology}

The prototype was evaluated using  \textbf{BIRD Bench} \citep{li2024can}, a widely recognized benchmark for testing natural language querying systems in relational databases. BIRD Bench presents a set of real-world queries that test the system's ability to interpret and translate natural language questions into SQL queries.

The BIRD Bench dataset includes a variety of queries, from simple single-table lookups to complex multi-table joins, aggregations, and domain-specific business logic. This allowed the prototype to be tested across different difficulty levels, providing a comprehensive view of its strengths and limitations.
   
The results are judged on the following criteria:

\begin{itemize}
\item[•] \textbf{SQL accuracy}: Whether the SQL query generated is syntactically correct and executable.
\item[•] \textbf{Data correctness}: Whether the retrieved data matches the expected results.
\end{itemize}

The other benefits of the approach such as the natural language response quality cannot be judged by the BIRD Bench.

\subsection{Initial Results}

\begin{figure*}

 \begin{center}
 \includegraphics[width=1.\linewidth]{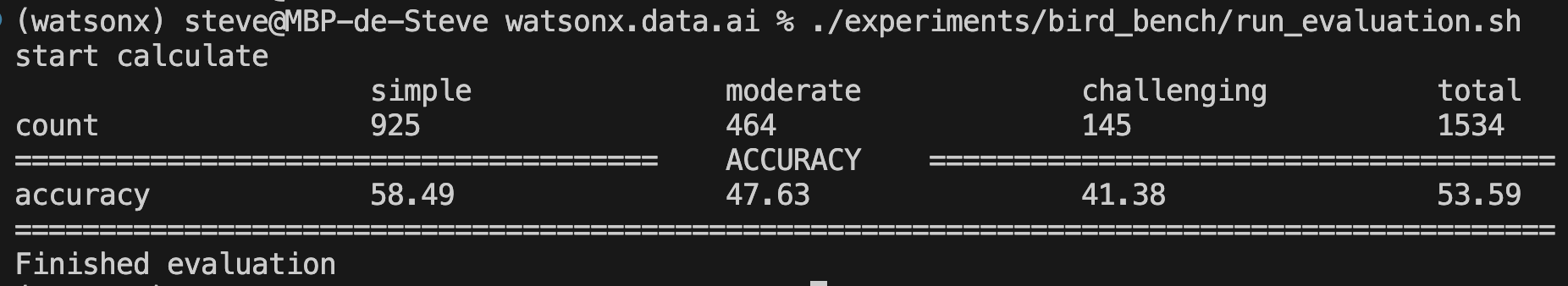}\caption{\label{initial_results} Bird Dev Bench Initial Results} 
 \end{center}
 
 \end{figure*}

The initial evaluation using the BIRD Bench yielded promising results (see \ref{initial_results}). The prototype demonstrated strong capabilities in generating correct SQL queries and returning accurate responses. The key findings include:

\begin{itemize}
\item[•] \textbf{Accuracy of Generated SQL}: Over \textit{50\% of the queries} were correctly translated into SQL on the first attempt (\textbf{more than 58\% for simple queries}). This is a significant achievement given the complexity of many of the benchmark queries, which often involve multi-table joins and some, even if few, domain-specific rules.
\item[•] \textbf{Handling of Business Rules}: The integration of business rules via the vector database proved to be effective in ensuring semantic correctness. Queries that involved domain-specific constraints, such as special calculations or relationships between business terms, were handled with high accuracy.
\item[•] \textbf{Improved Performance with Iteration}: With the iterative refinement process, the success rate increased significantly.
\item[•] \textbf{Natural Language Response Quality}: Even if this is not a direct output of the Bird Bench, the system’s ability to generate clear and well-formed natural language responses was evaluated on a dozen of nontechnical users. Users found the responses to be coherent and easy to understand, even for complex queries. The quality of responses was rated as \textbf{excellent} for 90\% of the test cases.

\end{itemize}

\subsection{Limitations and Areas for Improvement}

While the initial results are encouraging, there are several areas where the prototype can be improved:

\begin{itemize}
\item[•] \textbf{Handling of Edge Cases}: The system occasionally struggled with queries involving ambiguous language or highly specialized domain knowledge not captured in the vectorized business rules. 
\item[•] \textbf{Execution Time}: The iterative refinement process, while effective, increased the time required to generate correct responses for more complex queries. Optimizing the efficiency of the refinement process and minimizing the number of iterations required will be important for scaling the solution to real-world applications.
\item[•] \textbf{Scalability}: Although the prototype performed well on the BIRD Bench dataset, scaling it to larger datasets and more complex database structures (e.g., databases with hundreds of tables and intricate relationships) remains a challenge. Future evaluations will focus on stress-testing the system with larger, more complex datasets.
\end{itemize}

The prototyped system demonstrates the feasibility and potential of using generative AI and vector database technology to enable natural language querying of relational databases. The evaluation against BIRD Bench shows that the system is capable of translating natural language queries into accurate SQL queries, applying complex business rules, and returning user-friendly natural language responses. While there are areas for improvement, the system’s ability to handle business rules, generate coherent responses, and refine its outputs through iteration suggests that it is a solid foundation for future development.

\section{Conclusion}

The proposed method for natural language querying of relational databases represents a significant leap forward in making data access more intuitive and accessible for non-expert users. By leveraging the power of large language models, vector databases, and advanced AI-driven orchestration, the system breaks down the barriers of traditional SQL querying, allowing users to interact with complex data structures in a conversational and seamless way. Through syntactic and semantic validation of SQL queries, coupled with automatic response generation in natural language, this approach ensures accuracy, relevance, and user satisfaction.

The prototyping and evaluation of the method on IBM’s watsonx.data and watsonx.ai platforms demonstrated its potential to revolutionize data querying. With over 50\% of correct results in early tests (with more than 90\% of natural language responses rated \textbf{excellent} by nontechnical users), this approach shows promise in becoming a reliable system for enterprise-grade data access, significantly reducing the technical knowledge gap. By utilizing state-of-the-art technologies like Milvus for vector storage and LLMs like LLama 3 and Mixtral, the system bridges the gap between unstructured queries and structured database management with an impressive degree of accuracy and scalability.

The future of natural language querying is filled with exciting opportunities. As LLMs models become more sophisticated and the integration between data and AI platforms deepens, several promising avenues can be explored:

\begin{itemize}
\item[•] \textbf{Adaptation and Customization}: It may be interesting to tailor some LLMs models to specific industries and their business rules in order to further enhance query accuracy.
\item[•] \textbf{Enhanced Business Rule Handling}: Future iterations could introduce dynamic business rule generation, where AI not only applies pre-defined rules but also learns to infer new business rules based on patterns and trends in the data, and user journey. This would ensure even more accurate and contextually aware responses.
\item[•] \textbf{Expansion to NoSQL and Hybrid Databases}: While the current solution focuses on relational databases, extending its capabilities to handle NoSQL databases or hybrid data models would enable even more flexibility and power in handling diverse data ecosystems.
\end{itemize}

\bibliographystyle{elsarticle-num}
\bibliography{references}

\end{document}